\documentclass[12pt]{iopart}

\usepackage{iopams}  
\usepackage{amsmath}
\usepackage{graphicx}
\begin{document}

\title[]{Assessing the Suitability of the Langevin Equation for Analyzing Measured Data Through Downsampling}

\author{Pyei Phyo Lin\textsuperscript{1}, Matthias W\"achter\textsuperscript{1}, Joachim Peinke\textsuperscript{1}, \\ M. Reza Rahimi Tabar\textsuperscript{3,2,1}}

\address{\textsuperscript{1}ForWind, Institute of Physics, University of Oldenburg, Oldenburg,
Germany.}
\address{\textsuperscript{2}Theoretical Physics/Complex Systems, ICBM, University of
Oldenburg, Oldenburg, Germany.}
\address{\textsuperscript{3}Department of Physics, Sharif University of Technology, Tehran,
11155-9161 .}
\ead{pyei.phyo.lin@uol.de, matthias.waechter@uol.de, peinke@uol.de, tabar@uol.de}
\vspace{10pt}
\begin{indented}
\item[]November 2023
\end{indented}

\begin{abstract}
The measured time series from complex systems are renowned for their intricate stochastic behavior, characterized by random fluctuations stemming from external influences and nonlinear interactions. These fluctuations take diverse forms, ranging from continuous trajectories reminiscent of Brownian motion to noncontinuous trajectories featuring jump events. The Langevin equation serves as a powerful tool for generating stochasticity and capturing the complex behavior of measured data with continuous stochastic characteristics.
However, the traditional modeling framework of the Langevin equation falls short when it comes to capturing the presence of abrupt changes, particularly jumps, in trajectories that exhibit non-continuity. Such non-continuous changes pose a significant challenge for general processes and have profound implications for risk management. Moreover, the discrete nature of observed physical phenomena, measured with a finite sample rate, adds another layer of complexity. In such cases, data points often appear as a series of discontinuous jumps, even when the underlying trajectory is continuous. In this study, we present an analytical framework that goes beyond the limitations of the Langevin equation. Our approach effectively distinguishes between diffusive or Brownian-type trajectories and trajectories with jumps. By employing downsampling techniques, where we artificially lower the sample rate, we derive a set of measures and criteria to analyze the data and differentiate between diffusive and non-diffusive behaviors.  
To further demonstrate its versatility and practical applicability, we have applied our proposed method to real-world data in various scientific fields, turbulence, optical tweezers for trapped particles, neuroscience, renewable energy, and market price analysis.
\end{abstract}

\vspace{2pc}

\submitto{\JSTAT}
%\maketitle

\section{Introduction}\label{sec:intro}

In a complex world, empirical data ranging from natural science (e.g. physical, climatological, biological and ecological systems) to social sciences (e.g. infectious diseases spread, finance and economic) can be characterized by the stochastic dynamics. The classical method to approach such systems is to approximate the data with the Langevin equation or diffusion process \cite{Chandrasekhar1943, Friedrich2011, Peinke2019}. The Langevin equation is a stochastic differential equation which is driven by Gaussian white noise generating the continuous (Brownian type) stochastic process \cite{Gardiner1985, Tabar2019}. Recent studies have highlighted the presence of jumps, which are discontinuous or non-Brownian events, in complex systems. These jumps can have significant implications for risk management and have received considerable attention in recent years \cite{Hanson2007}. 
Jumps introduce higher uncertainties in the stochastic features of the underlying system and contribute to the non-Gaussian characteristics of the increment statistics in short-time scales \cite{Anvari2016a}. 

In the presence of discontinuous events, the traditional modeling framework of the Langevin equation, which generates only continuous trajectories, is inadequate for capturing these phenomena \cite{Tabar2019}. To address this limitation, an effective approximation is to introduce an additional noise term known as jump-noise in the Langevin equation with distributed jumps in time and amplitude. This gives rise to jump-diffusion processes \cite{Anvari2016a, Hanson2007, Tabar2019, Lin2022, Lin2023}. In this framework, the jump noise represents the discontinuous paths within the diffusion process. 

The nature of experimental data is such that it is often measured with limited sampling time or spatial resolution. As a result, the recorded trajectory may appear discontinuous, despite the underlying process being continuous. Furthermore, certain real-world datasets, such as those related to stock market indices, glaciers, and sea levels, exhibit distinct trajectories. Here, we present a powerful method to differentiate between diffusive and non-diffusive behavior in empirical data by using downsampling. This method is applicable to datasets with finite sampling intervals and those that display unique trajectories.

This paper is organized as follows. Firstly, we present the criteria to distinguish the diffusive and non-diffusive (jumpy) nature of the simulated stochastic processes with different integration time steps. Then, we discuss the applicability of these criteria to analyze the simulated data with a finite sampling time by mean of downsampling and their behavior at different downsampling time scales. Finally, we test these criteria on the real-world experimental data. All derivations and proofs are presented in appendices.

\section{Diffusive and non-diffusive behaviors}\label{sec:diffusive}

A given time series typically consists of three independent time scales: the sampling interval $\tau_s$, the correlation timescale $T_{\rm C}$, and the average timescale between jumps $T_{\rm J}$ (if present). The timescales  $T_{\rm C}$ and $T_{\rm J}$ can be estimated using Kramers–Moyal (KM) coefficients of order one, four and six, see below \cite{Nikakhtar2023}. For time series $x(t)$, KM coefficients of order $l$, $K^{(n)}({x},t)$ are given by
\begin{equation} \label{eq:KMcoeff}
	K^{(n)}(x, t) = \lim_{\tau_s\rightarrow 0} \frac{M^{(n)}(x, t, \tau_s)}{\tau_s}
\end{equation} 
where $M^{(n)}(x, t, \tau_s)$ are KM conditional moments, defined as
\begin{equation} \label{eq:condM}
	M^{(n)}(x, t, \tau_s) = \int dx' (x' - x)^n p(x', t + \tau_s | x, t) ~.
\end{equation}
In definition (\ref{eq:KMcoeff}), we omitted the $n!$ term in the denominator to simplify the expressions that allowing us to distinguish between diffusive and non-diffusive processes.
  
For a diffusive process that generates a continuous trajectory like Brownian motion \cite{Tabar2019}, the dynamics of a state variable $x(t)$ can be described by a Langevin equation that includes both deterministic and stochastic contributions, given by \cite{Risken1984}: 
\begin{equation}
\mathrm{d}x = D^{(1)}({x},t) \, \mathrm{d}t + \sqrt{D^{(2)}({x},t)} \, \mathrm{d}W_t ~,
\label{eq:diffusion}
\end{equation}
%}
where $D^{(1)}(x,t)$ (drift) and $D^{(2)}(x,t)$ (diffusion coefficient) are given functions of the state variable $x$ and ``time'' $t$. They are expressed in terms of the first and second KM coefficients as $D^{(1)}(x,t) = K^{(1)}(x,t)$ and $D^{(2)}(x,t) = K^{(2)}(x,t)$. All other KM coefficients are zero, i.e. $K^{(n)}(x,t) = 0$ for $n \geq 3$, for a diffusive process \cite{Pawula1967}. In Eq.~(\ref{eq:diffusion}), the increment of Wiener process is defined by $dW_t = \eta(t)dt$ where $\eta(t)$ represents a zero-mean Gaussian white noise with unit intensity, satisfying $\langle \eta(t) \eta(t^\prime) \rangle = \delta(t-t^\prime)$.

In practice, verifying the condition $K^{(n)}(x,t) = 0$ for $n \geq 3$ is not straightforward, as it requires knowledge of the expansions of KM conditional moments $M^{(n)}(x, t,\tau_s)$ in terms of the sampling time $\tau_s$. By employing the expansion of KM conditional moments, it has been demonstrated that the necessary condition for a process to belong to the diffusive category is $M^{(4)}(x, t,\tau_s) \simeq 3 (M^{(2)}(x, t, \tau_s))^2$ \cite{Lehnertz2018}. In an alternative representation, we define $\Theta (x,t,\tau_s)$ as:

\begin{equation}
\Theta (x, t, \tau_s) = 3 \frac{(M^{(2)}(x, t, \tau_s))^2}{M^{(4)}(x, t, \tau_s)} ~.
\label{eq:Theta}
\end{equation}

If $\Theta (x, t, \tau_s) \approx 1$, it indicates that the data fulfills the criterion for belonging to the diffusive processes. This criterion has been examined for various linear and nonlinear drift and diffusion coefficients, and we have observed $\Theta (x, t, \tau_s) \approx 1$; for more details, refer to %Appendix~\ref{app:simNLMN}.
\ref{app:simNLMN}.

For a given time series that $ \Theta (x, t, \tau_s)$ has not satisfying $\Theta (x, t, \tau_s) \approx 1$, one can approximate its dynamics with a jump-diffusion process. This can simply be done by adding to Langevin equations an additional Poisson distributed jump term (jump-diffusion process), therefore one can write, 

\begin{equation}
\mathrm{d}x = D^{(1)}({x},t) \, \mathrm{d}t + \sqrt{D^{(2)}({x},t)} \, \mathrm{d}W_t + \xi \, \mathrm{d}J_t ~.
\label{eq:jumpDiffusion}
\end{equation}

In this model, we consider the noise $\xi \sim \mathcal{N}(0, \sigma_\xi^2)$, which represents the size of jumps and follows a normal distribution with variance $\sigma_\xi^2$. The jump amplitude is also defined as, $\sigma_\xi^2$ and can be state and time-dependent. Additionally, the Poisson distributed jump process $J_t \sim P(\lambda t)$ is
present, which is a zero-one jump process with a jump rate of $\lambda(x,t)$.
Indeed, the final term in Eq.~(\ref{eq:jumpDiffusion}) simulates abrupt jumps in the trajectories of $x(t)$ with varying amplitudes. In the jump-diffusion dynamical equation (\ref{eq:jumpDiffusion}), we assume that jump events are rare and can be modeled via a Poisson process.

Two unknown functions, jump rate $\lambda(x,t)$ and jump amplitude, $\sigma_\xi^2$ can be found using the relations \cite{Anvari2016a, Tabar2019},
\begin{gather}
\sigma_{\xi}^{2}(x,t) = \frac{K^{(6)}(x,t)}{5 K^{(4)}(x,t)}, \hskip 0.5cm
\lambda(x,t) = \frac{K^{(4)}(x,t)}{3 \sigma_{\xi}^{4}(x,t)} ~.
\label{eq:jumpParameters}
\end{gather}

For jump-diffusion processes described by Eq.~(\ref{eq:jumpDiffusion}), expansion of KM conditional moments can be done and one can define a function $Q(x, t, \tau_s)$ (so-called $Q$-criterion, \cite{Lehnertz2018}) as (see %Appendix~\ref{app:condMJumpDiff}),
\ref{app:condMJumpDiff}),
\begin{equation}
 Q(x, t, \tau_s) =  \frac{M^{(6)}(x, t, \tau_s)}{5 M^{(4)}(x, t, \tau_s)} ~ . 
\label{eq:Q}
\end{equation}
The function $Q(x, t, \tau_s)$ serves as a criterion for determining if a jump is present in a given time series. 

For small values of $\tau_s$ in the functions $\Theta(x, t, \tau_s)$ and $Q(x, t, \tau_s)$, which are obtained from second, fourth, and sixth-order KM conditional moments, we observe distinct behavior for three types of time series (see %Appendix~\ref{app:condMexpan}), as 
\ref{app:condMexpan}), as 

\begin{equation}
\Theta(x, t, \tau_s) \approx 
\begin{cases}
1, &\text{diffusive}\\
\frac{(D^{(2)}(x, t) + \lambda(x, t) \sigma_\xi^2(x,t))^2}{\lambda(x, t) \sigma_\xi^4(x, t)} \tau_s, &\text{jumpy}\\
1, &\text{white noise at } x = 0, ~
\end{cases}
\label{eq:Thetadiffjump}
\end{equation}
\vskip 0.2cm
and
\vskip 0.2cm

\begin{equation}
Q(x, t, \tau_s) \approx 
\begin{cases}
D^{(2)}(x, t) \, \tau_s, &\text{diffusive},\\
\sigma_{\xi}^{2} (x, t), &\text{jumpy},\\
\sigma_\eta^2, &\text{white noise at}, ~ x = 0.
\end{cases}
\label{eq:Qdiffjump}
\end{equation} 

\vskip 1cm
Here, the behaviors of $\Theta(x, t, \tau_s)$ and $Q(x, t, \tau_s)$ for various processes, including diffusive (Langevin dynamics), jump-diffusion processes and pure white noise are given. It is worth noting that the requirement for $x=0$ is needed only in the case of Gaussian white noise. For all other situations, the aforementioned relations hold for any value of $x$. Also $\sigma_\eta^2$ is the variance of white noise.

The function $\Theta(x, t, \tau_s)$ exhibits a linear behavior with respect to $\tau_s$ for jump-diffusion and pure jump processes, while it remains unity for diffusive and white noise trajectories (at $x=0$). Simultaneously, the function $Q(x, t, \tau_s)$ takes on a constant value of $\sigma_{\xi}^{2}(x,t)$ for small $\tau_s$ in the case of jump-diffusion processes. However, for diffusive processes, it displays a linear relationship with $\tau_s$, specifically given by $D^{(2)}(x, t) \tau_s$.

%In the Appendix~\ref{app:QSimDerivation}, 
In the \ref{app:QSimDerivation}, we have conducted numerical integration of corresponding Langevin and jump-diffusion dynamics with various different values of $\tau_s$. This allowed us to verify the behaviors of $\Theta(x, t, \tau_s)$ and $Q(x, t, \tau_s)$ in relation to $\tau_s$ for diffusive processes, jump-diffusion processes and pure uncorrelated (white) noise. Therefore, by utilizing the $\Theta$- and $Q$-criterion, we can effectively distinguish whether a stochastic time series ``generated'' from certain dynamical equations originates from a diffusive (continuous) process or a jumpy (discontinuous) process \cite{Lehnertz2018}.

The results presented above for $\Theta(x, t, \tau_s)$ and $Q(x, t, \tau_s)$ are derived from the expansion of Kramers–Moyal coefficients for Langevin and jump-diffusion dynamical equations with varying time steps, $\tau_s$. To make them applicable to real-world time series, especially those displaying unique trajectories, their expressions require modification.

In the following, we demonstrate that the obtained behaviors of $Q(x, t, \tau)$ hold true for time scales $\tau_s$ that are shorter than both the correlation timescale $T_{\rm C}$ of the time series and the average timescale between jumps $T_{\rm J}$. However, when dealing with a real-world time series sampled with $\tau_s$, it is important to note that we typically lack direct access to the ground truth values of $T_{\rm C}$ and especially $T_{\rm J}$. Consequently, assessing the behaviors of $\Theta(x, t, \tau_s)$ and $Q(x, t, \tau)$ can potentially lead to misleading conclusions. Therefore, in this study, we aim to investigate this matter more deeply to gain a thorough understanding of the implications and limitations associated with using these criteria in real-world scenarios.

Based on a given dynamical equation, it is possible to generate a set of stochastic processes with different time scales $T_{\rm C}$ and $T_{\rm J}$ in order to check the $\Theta$- and $Q$-criteria for the corresponding time series for different $\tau_s$. In order to verify the these criteria using synthetic data, we conducted numerical integration of a Langevin equation with specific parameters. The drift coefficient is given by $D^{(1)}(x,t)=-\gamma x$, where $\gamma = 1~{\rm s^{-1}}$. The diffusion coefficient is denoted as $D^{(2)}(x,t)=D$, with $D = 1~{\rm s^{-1}}$. Additionally, we considered an additional jump term (jump-diffusion process), characterized by a jump rate of $\lambda = 100~{\rm s^{-1}}$ and a variance of $\sigma_\xi^2 = 0.01$. In the cases involving pure noisy data, we employed normal Gaussian white noise with a variance of $\sigma_\eta^2 = 1$. The chosen parameters in the Langevin and jump-diffusion dynamics yield two timescales. The correlation timescale is given by $T_{\rm C} = \frac{1}{\gamma} = 1~\mathrm{s}$, and the average timescale between jumps is denoted as $T_{\rm J}=\frac{1}{\lambda} = 0.01~\mathrm{s}$.

\begin{figure*}[t]
\includegraphics[width = \textwidth]{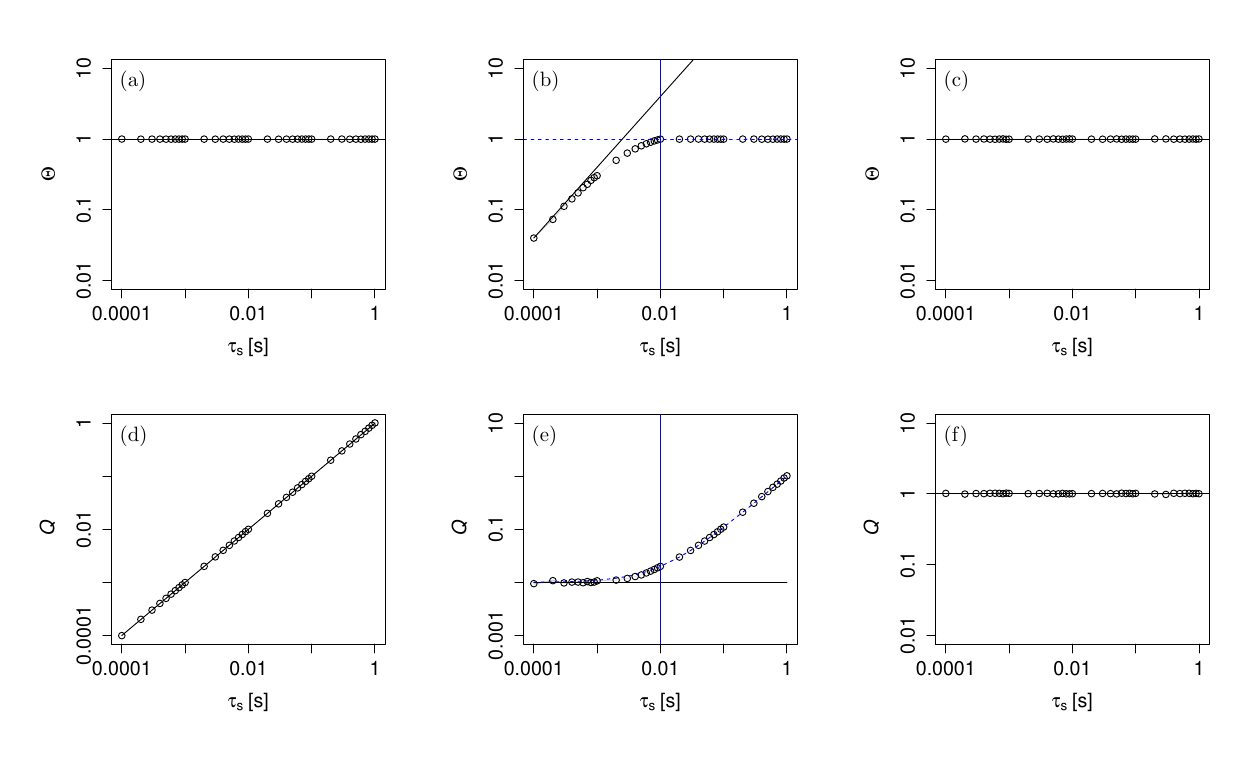} 
\caption{\label{fig:ThetaQdiffjump_sim} $\Theta = \Theta(x = 0, \tau = \tau_{\rm s})$ and $Q = Q(x = 0, \tau = \tau_{\rm s})$ are calculated for diffusion process, (a) and (d), jump-diffusion process, (b) and (e), and standard normal Gaussian white noise, (c) and (f), are evaluated for each $\tau_{\rm s}$. For the evaluation, OU processes are generated for different $\tau_{\rm s}$ with $\gamma = 1\,{\rm s^{-1}}$, $D = 1\,{\rm s^{-1}}$ and also with the additional jump terms with $\lambda = 100\,{\rm s^{-1}}$ and $\sigma_\xi^2 = 0.01$. The theoretical predictions according to Eqs.~(\ref{eq:Thetadiffjump}) and (\ref{eq:Qdiffjump}) are presented with black solid lines. The blue vertical line is the timescale $T_{\rm J} = \frac{1}{\lambda}$. Standard error of the mean are shown with shadows, which is smaller than the symbol size in these cases.}
\end{figure*}

From the simulated data with different time step $\tau_s$-values, we estimate $\Theta(x,\tau = \tau_s)$ and $Q(x,\tau = \tau_s)$ as shown in Fig.~\ref{fig:ThetaQdiffjump_sim}. 
Theoretical predictions as described in Eqs.~(\ref{eq:Qdiffjump}) and (\ref{eq:Thetadiffjump}) are also plotted with the solid lines. As shown in Fig.~\ref{fig:ThetaQdiffjump_sim}, for jump-diffusion process, Eq.~(\ref{eq:Qdiffjump}) is valid only for $\tau_s < \frac{1}{\lambda} = T_{\rm J}$. For small $\tau_s$ as shown in Fig.~\ref{fig:ThetaQdiffjump_sim}, all those estimates are rather close to the preset values and emphasize the accuracy of the predictions of $\Theta$ and $Q$-criteria for four types of the synthetic data. The subplot (ii) in Fig.~\ref{fig:ThetaQdiffjump_sim} exhibit a notable departure from the expected theoretical prediction of the Kramers-Moyal conditional moments' expansion for the jump-diffusion process, specifically observed in the case of $\tau_{\rm s} > T_{\rm J}$. To conclude this part, we find that the behaviors reported in Eqs.~(\ref{eq:Thetadiffjump}) and (\ref{eq:Qdiffjump}) are valid only for $\tau_{\rm s} < T_{\rm J}$ \text{and} $T_{\rm C}$. It is essential to consider these constraints when applying the criteria to real-world time series data.

\section{Downsampling of the data}\label{sec:downsampling}

To analyze data that is originally sampled with a finite time step $\tau_s$ and data with unique trajectories, the only practical approach to study diffusive and non-diffusive behaviors is downsampling. 
When downsampling data, we encounter situations where the sampling interval $\tau_{\rm ds}$ (which belongs to the set $\in{[\tau_s, 2\tau_s, \ldots]}$) can be larger than both the timescales $T_{\rm J}$ and $T_{\rm C}$, or it may lie between these two timescales. 

Initially, we focus on diffusive processes that are governed by Langevin equations. %In the Appendix~\ref{app:downsampling}, 
In the \ref{app:downsampling}, we provide a proof that downsampling time series generated from a Langevin equation does not alter the behavior of $\Theta (x, t, \tau_{\rm ds})$. Consequently, diffusive processes retain their diffusive nature even after downsampling. The key concept is that the Wiener process for the downsampled time series remain Wiener process, with the only difference being the change of $\tau_s$ to $\tau_{\rm ds}$.

However, downsampling behaves differently for processes that exhibit jump discontinuities. 
Let us consider the case where time scales have the property $T_{\rm J} < T_{\rm C}$. To this end, we note that on average, there are $\lambda \tau_{\rm ds}$ jump events within each downsampled step of $\tau_{\rm ds}$. Consequently, within each step, we observe multiple jumps with different amplitudes. The average jump amplitude for the downsampled time interval $\tau_{\rm ds}$ is given by:
\begin{equation}
\sigma_{\xi,{\rm eff}}^{2} = \sum_{i = 1}^{\lambda \tau_{\rm ds}} \sigma_{\xi}^{2} = \sigma_{\xi}^{2} \, \lambda  \tau_{\rm ds} ,
\label{eq:sig2downsampling}
\end{equation} 

where $\lambda$ represents the jump rate of the original time series (before downsampling). From Eq.~(\ref{eq:sig2downsampling}), we can observe that $\sigma_{\xi,{\rm eff}}^{2}$ now exhibits a linear dependence on $\tau_{\rm ds}$. As a result, it becomes challenging to distinguish the jump discontinuity process from the diffusive process using the $Q$-criterion presented in Eq.~(\ref{eq:Qdiffjump}), which we have shown that was applicable for small $\tau_{\rm s}$.

Similar to jump events, the Gaussian noise $\eta$ in the diffusion part of the jump-diffusion process also accumulates over each time step within the downsampled interval $\tau_{\rm ds}$. This results in an effective variance given by $\sigma_{\eta,{\rm eff}}^{2} = \sum_{i = 1}^{N_\tau} \sigma_{\eta}^{2} = N_\tau = \frac{\tau_{\rm ds}}{\tau_{\rm s}}$, where $\sigma_{\eta}^2 = 1$ represents the unit variance of $\eta$, and $\tau_{\rm s}$ is the sampling time step of the original data. As a result, the effective diffusion coefficient $D^{(2)}_{\rm eff}(x,t)$ satisfies the relation $D^{(2)}_{\rm eff}(x,t) \tau_{\rm ds} = D^{(2)} \tau_{\rm s} N_\tau$, which implies that the effective diffusion coefficient remains the same as the original diffusion coefficient ($D^{(2)}_{\rm eff}(x,t) = D^{(2)}(x,t)$). A detailed proof can be found %in the Appendix~\ref{app:downsampling}. 
in the \ref{app:downsampling}. Thus, the effective function $Q$ for the jump-diffusion in the range of $T_{\rm J} < \tau_{ds} < T_{\rm C}$ can be written as
\begin{equation}
Q_{\rm eff} = D^{(2)} ~ \tau_{\rm ds} + \lambda \sigma_\xi^2 ~ \tau_{\rm ds} ~.
\label{eq:Qdownsampling}
\end{equation}
Therefore, in Eq.~(\ref{eq:Qdiffjump}), the condition for jumpy time series changes from a constant $\sigma_\xi^2$ to a linear relation with $\tau_{\rm ds}$, given by $(D^{(2)} + \lambda \sigma_\xi^2 ) \tau_{\rm ds}$. If $\tau_{\rm ds} > T_{\rm C}$, the downsampled data will resemble white noise. In this scenario, the $Q$-criterion will have a value equal to the variance of the resulting time series. We note also that in the case of $T_{\rm C} < T_{\rm J}$, Eqs.~(\ref{eq:Thetadiffjump}) and (\ref{eq:Qdiffjump}) are valid for $\tau_{\rm ds} < T_{\rm C}$.

In order to check the aforementioned expressions, i.e. Eqs.~(\ref{eq:Thetadiffjump}), (\ref{eq:Qdiffjump}) and (\ref{eq:Qdownsampling}) numerically, we generate one sample each of OU process and OU jump-diffusion process with $\gamma = 1~{\rm s^{-1}}$, $D = 1~{\rm s^{-1}}$, $\lambda = 100~{\rm s^{-1}}$ and $\sigma_\xi^2 = 0.01$. For the case of $T_{\rm J} > T_{\rm C}$, we generate additional sample with $\gamma = 1000~{\rm s^{-1}}$, $D = 1000~{\rm s^{-1}}$, $\lambda = 100~{\rm s^{-1}}$ and $\sigma_\xi^2 = 10$. All the samples are simulated with the integrating time step $\tau_{\rm s} = 10^{-4}~{\rm s}$.

\begin{figure*}[t]
\includegraphics[width = \textwidth]{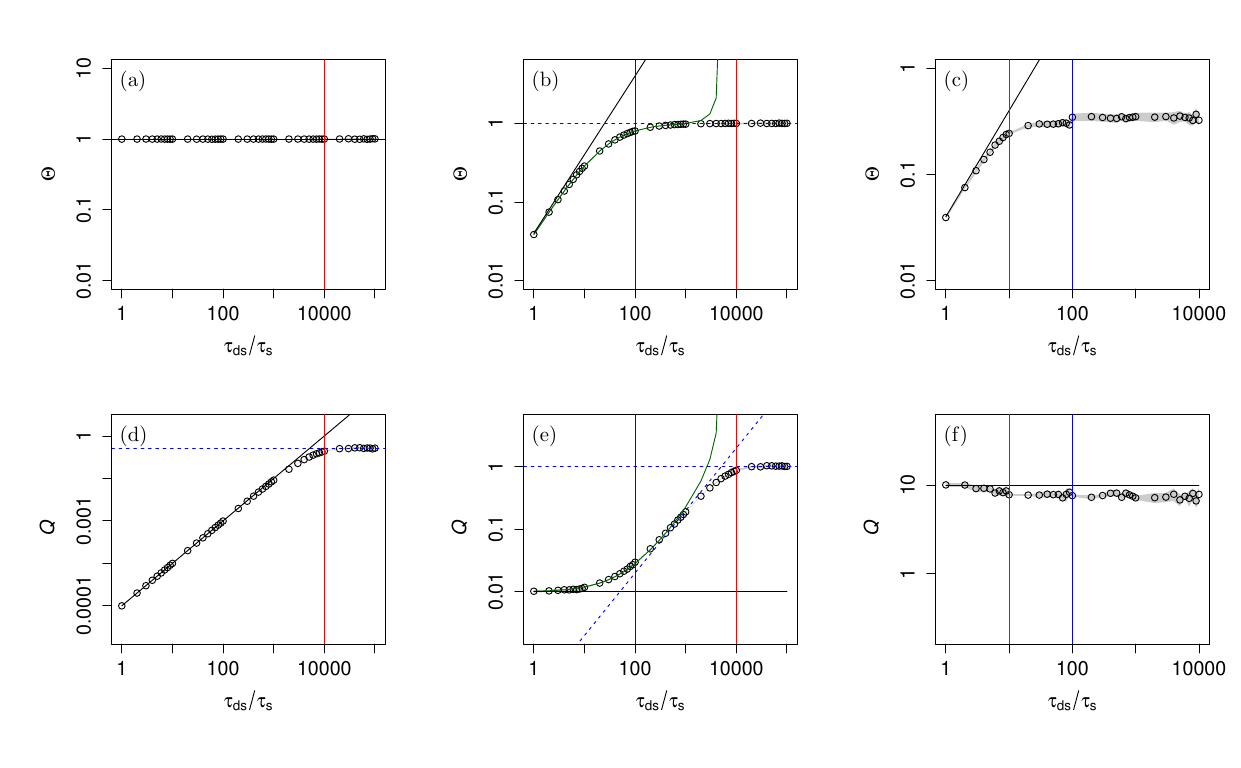} 
\caption{\label{fig:ThetaQdiffjump_ds} $\Theta = \Theta(x = 0, \tau = \tau_{\rm s})$ and $Q = Q(x = 0, \tau = \tau_{\rm s})$ are calculated for the downsampled OU (diffusion) process, (a) and (d), downsampled OU jump-diffusion process for $T_{\rm J} < T_{\rm C}$, (b) and (e), and $T_{\rm J} > T_{\rm C}$, (c) and (f), are evaluated for each downsampled period $\tau_{\rm s}$. The theoretical predictions according to Eqs.~(\ref{eq:Thetadiffjump}), (\ref{eq:Qdiffjump}) and (\ref{eq:Qdownsampling}) are presented with black solid lines. The blue vertical line is the timescale $T_{\rm J} = \frac{1}{\lambda}$ and the red vertical line $T_{\rm C} = \frac{1}{\gamma}$. Standard error of the mean are shown with shadows, which is smaller than the symbol size in these cases. The green solid lines in (b) and (e) the approximation up to $\mathcal{O}(\tau^4)$ of conditional moments while calculating $\Theta$ and $Q$ in Eqs.~(\ref{eq:Thetadiffjump}) and (\ref{eq:Qdiffjump}) as described %in Appendix~\ref{app:condMJumpDiff}.}
in \ref{app:condMJumpDiff}.}
\end{figure*}

The processes are realized with different time-lags $\tau_{\rm ds}$ using downsampling, and then the $\Theta$- and $Q$-criterion are calculated. The results are plotted in Fig.~\ref{fig:ThetaQdiffjump_ds}. The blue vertical line is the timescale $T_{\rm J} = \frac{1}{\lambda}$ and the red vertical line $T_{\rm C} = \frac{1}{\gamma}$. 

For diffusion process, we can observe the same diffusive behavior for $\tau_{\rm ds} < T_{\rm C} = \frac{1}{\gamma}$ as described in Eq.~(\ref{eq:Thetadiffjump}) and (\ref{eq:Qdiffjump}) shown by black solid lines in Fig.~\ref{fig:ThetaQdiffjump_ds}~(a) and (d). For downsampled time steps larger than correlation length, $\tau_{\rm ds} > T_{\rm C}$, the results of the criteria shows the behavior of uncorrelated Gaussian white noise in which  $\Theta = 1$ and $Q = \sigma_x^2$ where $\sigma_x^2 = 0.5$ of OU process $x$, shown by blue horizontal dashed in Fig.~\ref{fig:ThetaQdiffjump_ds}~(d) for this case. 

In the first case of jump-diffusion process where $T_{\rm J} < T_{\rm C}$ shown in Fig.~\ref{fig:ThetaQdiffjump_ds}~(b) and (e), we can see that the approximations in Eqs.~(\ref{eq:Thetadiffjump}) and (\ref{eq:Qdiffjump}) are still valid for $\tau_{\rm ds} \ll T_{\rm J}$. (If we consider calculating $\Theta$ and $Q$ in Eqs.~(\ref{eq:Thetadiffjump}) and (\ref{eq:Qdiffjump}) up to $\mathcal{O}(\tau^4)$ of conditional moments as described %in Appendix~\ref{app:condMJumpDiff}, 
in \ref{app:condMJumpDiff}, we can have a better approximation shown in solid green line.) In the range of $T_{\rm J} < \tau_{\rm ds} < T_{\rm C}$, the effective value of $Q$ becomes the sum of effective diffusion and jump terms described in Eq.~(\ref{eq:Qdownsampling}), and $\Theta$ also becomes unity after a few time steps in this range which apparently show the diffusive behavior. These are shown with the blue dashed lines. After $\tau_{\rm ds} > T_{\rm C}$, the downsampled process becomes uncorrelated with Gaussian distribution, where $\Theta = 1$ and $Q$ becomes the variance of the process $x$, $Q = \sigma_x^2 = 1$. Theoretically, the variance of the OU process with jump can be calculated with Eq.~(\ref{eq:OUjumpVar2}).

\begin{figure*}[t]
\centering
\includegraphics[width=\textwidth]{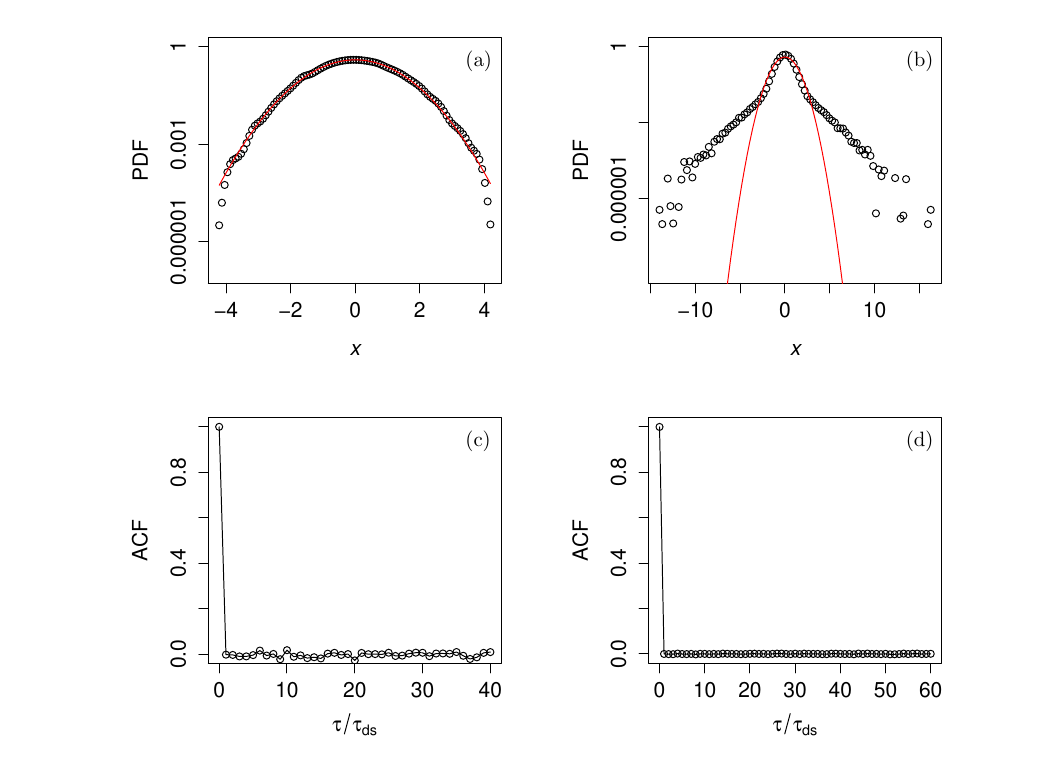}
\caption{\label{fig:PDFACFjumpDiff} The PDF and ACF of downsampled OU jump-diffusion process $x$ at $\tau_{\rm ds} > T_{\rm C}$ in the case of $T_{\rm J} < T_{\rm C}$, (a) and (c), and at $\tau_{\rm ds} > T_{\rm J}$ in the case of $T_{\rm J} > T_{\rm C}$, (b) and (d). The red solid line shows the PDF of Gaussian distribution with zero mean and the same variance $\sigma_x^2$. The PDF is plotted in semi-logarithmic scale.}
\end{figure*}

In the second case of jump-diffusion process where $T_{\rm J} > T_{\rm C}$ shown in Fig.~\ref{fig:ThetaQdiffjump_ds}~(c) and (f), we can see that the approximations in Eqs.~(\ref{eq:Thetadiffjump}) and (\ref{eq:Qdiffjump}) are still valid for $\tau_{\rm ds} \ll T_{\rm C}$ (shown by black solid lines). For $\tau_{\rm ds} > T_{\rm J}$, $Q$, $\Theta$ and $\Lambda$ become constant. In this range, the process becomes uncorrelated but not Gaussian distributed. It can be seen from the probability density function (PDF) and autocorrelation function (ACF) of process $x$ downsampled at $\tau_{\rm ds} > T_{\rm C}$ for first case and $\tau_{\rm ds} > T_{\rm J}$ for second of the aforementioned jump diffusion processes as shown in Fig.~\ref{fig:PDFACFjumpDiff}

\section{Real-world examples}\label{sec:realworld}

All the measured data $x(t)$ with the sampling time $\tau_{\rm s}$ are subtracted with their means and divided with their standard deviations, so that the normalized data have zero-means and standard deviations of one. The $Q$-, $\Theta$- and $\Lambda$-criterion are evaluated at $x = 0$ by downsampling of the given data with downsampling time step $\tau_{\rm ds}$. We take five different data sets (including the dynamics of trapped particles in optical tweezers, solar irradiance, turbulence, log return of DAX data and intracranial electroencephalographic (iEEG) time series of the brain) and separated into three categories.

For each time series sampled with timescale $\tau_s$, we apply the methods described %in the Appendix~\ref{app:condMexpan}
in the \ref{app:condMexpan} to estimate the drift, diffusion, jump rate, and jump amplitude. This involves analyzing the KM coefficients of order $1$, $4$, and $6$ to derive the values of $T_{\rm C}$ and $T_{\rm J}$ for the respective data sets. 

\begin{figure*}[t]
\vspace{-33pt} 
\centering
\includegraphics[width=0.7\textwidth]{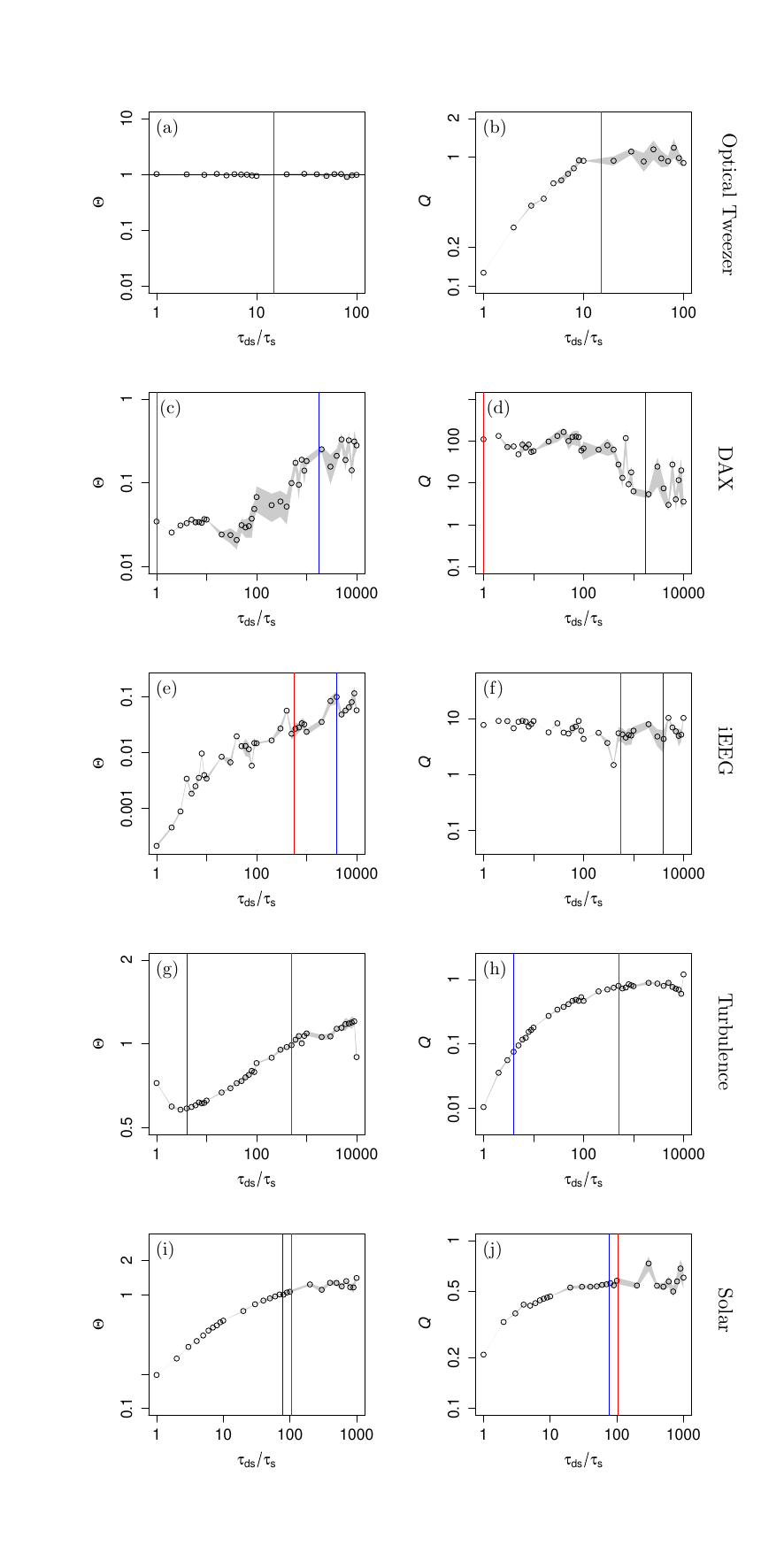} \vspace{-33pt} 
\caption{\label{fig:realdata} $\Theta$- and $Q$-criterion of downsampled displacement data of dielectric bead in the optical tweezers, (a) and (b), log return of DAX data, (c) and (d), iEEG brain data, (e) and (f), turbulence data, (g) and (h), and solar clear sky index data, (i) and (j), are plotted in double logarithmic scale. The vertical blue lines show the jump time scale $T_{\rm J} = \frac{1}{\lambda}$ and the vertical red lines the correlation time scale $T_{\rm C} = \frac{1}{\gamma}$. Standard errors of the mean are shown as gray-shaded background and some of them are smaller than the symbol size.}
\end{figure*}

In the first category, we analyze the data of measurements of the spatial positions of a dielectric bead (polystyrene, diameter-1${\rm \mu m}$, Bangs Laboratories Inc. USA) trapped in and optical tweezers \cite{Mousavi2017}. The results of $\Theta = \Theta(x = 0, \tau_{\rm ds})$ and $Q = Q(x = 0, \tau_{\rm ds})$ versus $\tau_{\rm ds}$, for a single measured data with duration $3$~s, and with sample rate $22$~kHz are plotted in Fig.~\ref{fig:realdata}~(a) and (b). 

The results from optical tweezers show the behavior of diffusion process. The correlation time $T_{\rm C} = \frac{1}{\gamma}$ is first estimated from the drift function, which is approximately $D^{(1)}(x) = -\gamma x$. Below the correlation timescale $\tau_{\rm ds} < T_{\rm C}$, $Q$ has the linear behavior and $\Theta = 1$. For $\tau_{\rm ds} > T_{\rm C}$, $Q = \sigma_x^2 = 1$ and $\Theta = 1$ also, which show the behavior of uncorrelated Gaussian white noise.

In the second category, we analyze DAX stock market index data \cite{Ghashghaie1996} and epileptic brain data \cite{Anvari2016a}. 
For the DAX stock market index, the log return is used for the analysis. The results of $\Theta = \Theta(x = 0, \tau_{\rm ds})$ and $Q = Q(x = 0, \tau_{\rm ds})$ verses $\tau_{\rm ds}$ are plotted in Fig.~\ref{fig:realdata}.~(c) and (d). The results of $\Theta = \Theta(x = 0, \tau_{\rm ds})$ and $Q = Q(x = 0, \tau_{\rm ds})$ verses $\tau_{\rm ds}$ for epileptic brain data are plotted in Fig.~\ref{fig:realdata}~(e) and (f). 

Since $\Theta$ are deviated from one, they do not fulfil the necessary conditions for the diffusion process. Here we evaluated again the correlation time $T_{\rm C} = \frac{1}{\gamma}$ obtained from the drift function, which is approximately $D^{(1)}(x) = -\gamma x$. We also estimated $T_{\rm J} = \frac{1}{\lambda}$ by Eq.~(\ref{eq:jumpParameters}). For small $\tau_{\rm ds}$, $Q$ shows the constant behavior while $\Theta$ is not unity and also show the non-constant behavior on $\tau$. Thus, one could consider the jump-diffusion process to model these dynamics \cite{Anvari2016a}.

In the third category, we look into the data of free jet low temperature helium turbulence \cite{Manshour2015} with Reynolds number $Re = 757,000$ and solar clear sky index \cite{Anvari2016b, Madanchi2017}. The results of $\Theta = \Theta(x = 0, \tau_{\rm ds})$ and $Q = Q(x = 0, \tau_{\rm ds})$ verses $\tau_{\rm ds}$ for the turbulence data are plotted in Fig.~\ref{fig:realdata}~(g) and (h). The results of $\Theta = \Theta(x = 0, \tau_{\rm ds})$ and $Q = Q(x = 0, \tau_{\rm ds})$ verses $\tau_{\rm ds}$ for the solar clear sky index are plotted in Fig.~\ref{fig:realdata}~(i) and (j). As $\Theta$ is deviated from unity, they do not fulfil the necessary conditions for the diffusion process. Similarly to previous data analysis, we evaluated again the correlation time $T_{\rm C} = \frac{1}{\gamma}$ where the drift function is defined by $D^{(1)}(x) = -\gamma x$. We also estimated $T_{\rm J} = \frac{1}{\lambda}$ by Eq.~(\ref{eq:jumpParameters}). In these two cases, $T_{\rm J} < T_{\rm C}$. For small $\tau_{\rm ds}$, both $Q$ and $\Theta$ show the nonlinear behavior on $\tau$. Thus, one could only conclude that they are not diffusive from the current study. Turbulence is known to have much more complex nature, and it is also not Markovian in time but is in scale \cite{Renner2001, Friedrich2011}. For the third category, we can only conclude that these time series do not belong to the class of diffusion processes.

\section{Conclusion}\label{sec:conclusion}

In many models describing the dynamics of natural measurements, it is common to assume a standard Gaussian white noise-driven Langevin equation to account for observed variability. However, it is not immediately clear whether this assumption holds true. Evaluating the validity of a Langevin equation to describe the data presents a significant challenge, especially when dealing with finite data points or processes that exhibit unique trajectories \cite{Manshour2015, Movahed2011, Madanchi2017}.

In this study, we have developed a set of criteria to assess the suitability of modeling the data using a Langevin equation. By recognizing the limitations of Langevin dynamics, alternative modeling approaches become relevant. These may include jump-diffusion stochastic dynamics \cite{Anvari2016a}, 
generalized Langevin equations~\cite{Kou2004, Lei2016}, fractional Klein-Kramers equations~\cite{Dieterich2008}, fractional diffusion processes~\cite{Kassel2022, Kassel2023} and L\'evy-driven Langevin dynamics \cite{Tabar2019}.

Revealing novel aspects of the data that cannot be captured by white-noise driven Langevin equations can illuminate the presence of abrupt changes in time series. These changes often arise from complex physical phenomena that extend beyond the scope of Langevin equations. Consequently, this exploration uncovers new phenomena, enhances our understanding of measured variability in diverse datasets, and opens up avenues for further research.

\clearpage

\appendix

\section{Expansion of conditional moments at finite time step}\label{app:condMexpan}

The conditional probability distribution of the process $x = x(t)$ which satisfies the Kramers-Moyal differential equation can be written as \cite{Lehnertz2018, Tabar2019},

\begin{equation}\label{eq:KMexp}
    \frac{\partial p(x,t | x^\prime,t^\prime)}{\partial t} =  \mathcal L_{\rm KM} ~ p(x,t | x^\prime,t^\prime)
\end{equation}

with initial condition $p(x,t | x^\prime,t) = \delta(x-x^\prime)$ and the Kramers-Moyal (KM) operator $\mathcal L_{\rm KM}$ is given by,

\begin{equation}
    \mathcal L_{\rm KM} =  \sum_{n=1}^{\infty} \frac{1}{n!}\left(-\frac{\partial}{\partial x}\right)^n ~ K^{(n)}(x,t) \quad.
    \label{eq:KMoperator}
\end{equation}
  
The formal solution of (\ref{eq:KMexp}) reads

\begin{equation}
    p(x,t+\tau|x^\prime,t)=  \exp\{\tau ~ \mathcal L_{\rm KM}\} ~ \delta(x-x^\prime) \quad .
    \label{eq:KMsolution}
\end{equation}

The $n^{th}$-order conditional moments $M^{(n)}(x,t,\tau)$ with finite $\tau$ can be written as,

\begin{eqnarray}
    M^{(n)}(x_i,t,\tau) &=& \int_{-\infty} ^{\infty} (x-x_i)^n \exp\{\tau ~ \mathcal L_{\rm KM}\}  ~ \delta(x-x_i) dx \cr \nonumber \\
    &=&  \exp\{\tau ~ \mathcal L_{\rm KM}^{\dagger} \} (x-x_i)^n \vert_{x=x_i}
\label{eq:condMtau}
\end{eqnarray}

where $ \mathcal L_{\rm KM}^{\dagger}$ is the adjoint operator of $\mathcal L_{\rm KM}$ and is given by,

\begin{equation}
    \mathcal L_{\rm KM}^{\dagger} =  \sum_{n=1}^{\infty} \frac{1}{n!} ~ K^{(n)}(x,t) \left(\frac{\partial}{\partial x}\right)^n.
\end{equation}

where $K^{(n)}(x,t) =  \lim_{\tau \to 0} \frac{1}{\tau} M^{(n)}(x,t,\tau)$ are the $n^{th}$-order Kramers-Moyal coefficients and $M^{(n)}(x,t,\tau)$  are given by Eq.~(\ref{eq:condMtau}). Next, the explicit cases of the diffusion and jump-diffusion processes are discussed.

\subsection{Conditional moments of diffusion and jump-diffusion processes}\label{app:condMJumpDiff}

\underline{Diffusion process}, in general, can be described by Langevin equation in Eq.~(\ref{eq:diffusion}). We can evaluate its conditional moments $M_{\rm d}^{(n)}(x,\tau)$ using Eq.~(\ref{eq:condMtau}). For diffusion process or Langevin equation, the adjoint KM operator $\mathcal L_{\rm KM}^{\dagger}$ becomes the adjoint Fokker-Planck (FP) operator $\mathcal L_{\rm FP}^{\dagger}$. For better readability, let the drift be $D^{(1)}(x,t) = K^{(1)}(x,t) = a$ and the diffusion $D^{(2)}(x,t) = K^{(2)}(x,t) = b^2$ and the adjoint FP-operator becomes

\begin{equation}
    \mathcal L_{\rm FP}^{\dagger} =  a~\frac{\partial}{\partial x} + \frac{1}{2}~ b^2~\frac{\partial^2}{\partial x^2}~.
    \label{eq:FPoperator}
\end{equation}

In this formulation, $a$ and $b$ can still be the function of $x$ and $t$. The second-, fourth- and sixth-order conditional moments up to first non-vanishing term can be derived from Eqs.~(\ref{eq:condMtau}) and (\ref{eq:FPoperator}) as follow \cite{Lehnertz2018}: 

\begin{eqnarray}
M_{\rm d}^{(2)}(x, \tau ) &=&   b^2 \tau  + \mathcal{O}(\tau^2), \cr \nonumber \\
M_{\rm d}^{(4)}(x, \tau ) &=& 3  b^4 \tau^2  + \mathcal{O}(\tau^3), \cr \nonumber \\
M_{\rm d}^{(6)}(x, \tau ) &=& 15 b^6 \tau^3 +  \mathcal{O}(\tau^4) .
\label{eq:condMdiff}
\end{eqnarray}

In order to distinguish from the jump-diffusion process which will be discussed next, %we used 
the subscript ``d'' is used here for the diffusion process.

\bigskip
\noindent
\underline{Ornstein-Uhlenbeck (OU) process} is considered here to be more concrete such that

\begin{equation}
\mathrm{d}x = -\gamma x ~ \mathrm{d}t + \sqrt{D} ~ \mathrm{d}W_t~.
\label{eq:OU}
\end{equation}

where $\gamma$ and $D$ are positive real constants. Expansion of $M_{\rm d}^{(n)}(x, t, \tau)$ of OU process for small $\tau$ up to $\mathcal{O} (\tau^3)$ (up to $\mathcal{O} (\tau^4)$ for sixth-order conditional moment) reads 

\begin{eqnarray}
M_{\rm d}^{(2)}(x, \tau ) &=&   D \tau + \left(\gamma^2 x^2 - \gamma D \right) \tau^2 + \mathcal{O}(\tau^3), \cr \nonumber \\
M_{\rm d}^{(4)}(x, \tau ) &=& 3 D^2 \tau^2  + \mathcal{O}(\tau^3), \cr \nonumber \\
M_{\rm d}^{(6)}(x, \tau ) &=& 15 D^3 \tau^3 +  \mathcal{O}(\tau^4) .
\label{eq:condMdiffOU}
\end{eqnarray}

\bigskip
\noindent
\underline{Jump-diffusion process}, in general, is described in Eq.~(\ref{eq:jumpDiffusion}). The KM-coefficients of such jump-diffusion are  

\begin{eqnarray}
    K^{(1)}(x, t) &=& D^{(1)}(x, t) \cr \nonumber \\
    K^{(2)}(x, t) &=& \left[ D^{(2)}(x, t) + \langle \xi^2 \rangle \lambda(x)\right] \cr \nonumber \\
    K^{(2n)}(x, t) &=& \langle \xi^{2n} \rangle \lambda(x), ~ ~\text{for} ~2n>2
\label{eq:KMcoefjump}
\end{eqnarray}

Therefore, the adjoint KM-operator $\mathcal L_{\rm KM}^{\dagger}$ for jump-diffusion process becomes

\begin{eqnarray}
    \mathcal L_{\rm KM}^{\dagger} &=&  \underbrace{ D^{(1)}(x, t)}_\textbf{A}  \frac{\partial}{\partial x}+
    \underbrace{\frac{[ D^{(2)}(x, t) + \langle \xi^2 \rangle \lambda(x)]}{2!}}_\textbf{B}  \frac{\partial^2}{\partial x ^2} \cr \nonumber \\
    &+& \underbrace{\frac{ \langle \xi^4 \rangle \lambda(x)}{4!} }_\textbf{C} \frac{\partial^4}{\partial x ^4}+
    \underbrace{  \frac{ \langle \xi^6 \rangle \lambda(x)}{6!} }_\textbf{D}   \frac{\partial^6}{\partial x ^6}
    \nonumber \\
    &+& \underbrace{\frac{ \langle \xi^8 \rangle \lambda(x)}{8!} }_\textbf{E}  \frac{\partial^8}{\partial x ^8}+
    \cdots \quad.
    \label{eq:KMjumpoperator}
\end{eqnarray}

Again for better readability, we use the substitution of the coefficients with $\textbf{A},\textbf{B},\textbf{C},\textbf{D},\textbf{E},\cdots$, which can still be the function of $x$ and $t$ in this formulation. With it, we can derive the conditional moments of the jump-diffusion equation Eq.~(\ref{eq:jumpDiffusion}) for second-, fourth and sixth- orders of the time interval $\tau$ using Eq.~(\ref{eq:condMtau}) and (\ref{eq:KMjumpoperator}) such that

\begin{eqnarray}
M_{\rm j}^{(2)}(x,\tau) &=& 2 \textbf{B} \tau + \mathcal{O}(\tau^2), \cr \nonumber \\
M_{\rm j}^{(4)}(x,\tau) &=& 4! \textbf{C} \tau + \mathcal{O}(\tau^2), \cr \nonumber \\
M_{\rm j}^{(6)}(x,\tau) &=& 6! \textbf{D} \tau + \mathcal{O}(\tau^2),
\label{eq:condMjump}
\end{eqnarray}

where the subscript ``j'' denotes the jump-diffusion process \cite{Lehnertz2018}. Here, they are derived up to first non-vanishing term. 

\bigskip
\noindent
\underline{OU process with an additional jump term} with constant jump rate and jump amplitude is now considered. It is a linear jump-diffusion process, and one finds,

\begin{equation}
\mathrm{d}x = -\gamma x ~ \mathrm{d}t + \sqrt{D} ~ \mathrm{d}W_t + \xi ~ \mathrm{d}J_t ~.
\label{eq:OUjump}
\end{equation}

Adding the jump term in OU process creates the discontinuities in the trajectory. 
For finite time interval $\tau$, the conditional moments $M_{\rm j}^{(n)}(x,\tau)$ of an OU jump-diffusion process can be determined up to $\mathcal{O}(\tau^4)$ using Eqs.~(\ref{eq:condMtau}) and (\ref{eq:KMjumpoperator}) as follows:  

\begin{eqnarray}
M_{\rm j}^{(2)}(x,\tau) &=&  (D + \lambda \sigma_\xi^2) \tau + \left(\gamma^2 x^2 - \gamma(D + \lambda \sigma_\xi^2)\right)\tau^2 \cr \nonumber \\ 
&+& \left(\frac{1}{3} \gamma^2 (D + \lambda \sigma_\xi^2) - \frac{1}{2} \gamma^3 x^2 \right)\tau^3 + \mathcal{O}(\tau^4), \cr \nonumber \\
M_{\rm j}^{(4)}(x,\tau) &=&  3 \lambda \sigma_\xi^4 \tau + 3\left((D + \lambda \sigma_\xi^2)^2 - 2 \gamma \lambda \sigma_\xi^4  \right)\tau^2 \cr \nonumber \\ 
&+& (6 x^2 \gamma^2 (D + \lambda \sigma_\xi^2) + 8 \gamma^2 \lambda \sigma_\xi^4 \cr \nonumber \\
&-& 6 \gamma (D + \lambda \sigma_\xi^2)^2)\tau^3 + \mathcal{O}(\tau^4),  \cr \nonumber \\
M_{\rm j}^{(6)}(x,\tau) &=& 15 \lambda \sigma_\xi^6 \tau + 45 \left(\lambda \sigma_\xi^4 (D + \lambda \sigma_\xi^2) -\gamma \lambda \sigma_\xi^6 \right) \tau^2 \cr \nonumber \\ %&+& \mathcal{O}(\tau^3),
&+&  ( 45 \gamma^2 \lambda \sigma_\xi^4 x^2 + 15(D + \lambda \sigma_\xi^2)^3 \cr \nonumber \\
&+& 540 \gamma^2 \lambda \sigma_\xi^6 - 810 \gamma \lambda \sigma_\xi^4 (D + \lambda \sigma_\xi^2)) \tau^3 + \mathcal{O}(\tau^4), \cr \nonumber \\
\label{eq:condMjumpOU}
\end{eqnarray}

\subsection{Conditional moments of gaussian white noise}\label{app:condMGauss}

For a Gaussian distributed white noise with zero mean and variance $\sigma_\eta^2$, $x = \eta \sim N(0,\sigma_\eta^2)$, the $n^{\rm th}$ order conditional moments can be calculated as

\begin{eqnarray}
    M_{\rm g}^{(n)}(x, \tau ) &=&  \left\langle \left( x(t+\tau) - x(t) \right)^n \vert_{x(t) = x} \right\rangle, \cr \nonumber \\
    &=& \int_{-\infty}^{\infty} \left( x^\prime - x \right)^n \cdot p(x^\prime, t^\prime \vert x,t) \cdot {\rm d}x^\prime, \cr \nonumber \\
    &=& \int_{-\infty}^{\infty} \left( x^\prime - x \right)^n \cdot p(x^\prime, t^\prime) \cdot {\rm d}x^\prime .
\label{eq:condMG}
\end{eqnarray}

For $x = 0$ and $p(x^\prime, t^\prime) = p(x^\prime) = \frac{1}{\sqrt{2 \pi \sigma_\eta^2}} \exp \left( -\frac{{x^\prime}^{2}}{2 \sigma_\eta^2} \right)$, we can use Wick's theorem \cite{Isserlis1916, Wick1950} to solve integral in Eq.~(\ref{eq:condMG}). The $(2n + 1)^\textrm{th}$ and the $2n^\textrm{th}$ conditional moments of Gaussian white noise become  

\begin{eqnarray}
    M_{\rm g}^{(2n+1)}(x=0, \tau ) &=& 0, \cr \nonumber \\
    M_{\rm g}^{(2n)}(x=0, \tau ) &=& \frac{(2n)!}{2^n n!} \sigma_\eta^{2n} ~,
\label{eq:condMGr}
\end{eqnarray}

where $\sigma_\eta^2$ is its variance.   

\subsection{$Q$-criterion of jump-diffusion process at time scale $\tau > T_{\rm J}$}\label{app:QSimDerivation}

As a first criterion, the $Q$-criterion is introduced in \cite{Lehnertz2018} to distinguish whether given synthetic data simulated with different integration time step $\tau_{\rm s}$ are diffusive or jumpy. Here, this criterion is also applied to the uncorrelated Gaussian white noise.

For a small $\tau$, we obtain the function $Q(x,\tau)$ as described in Eq.~(\ref{eq:Qdiffjump}) considering up to the first non-vanishing order term \cite{Lehnertz2018} which is derived from the fourth- and sixth-order conditional moments. 
%Note that $x=0$ is required only for Gaussian white noise, otherwise the relations hold for all $x$. 

For diffusive process, case~(i), the function $Q$ has a linear relationship with $\tau$ while it is constant in first order approximation in $\tau$ for jump-diffusion process, case~(ii). Additionally, the function $Q$ is also constant for the Gaussian white noise, case~(iii) which can be derived by direct computations of conditional moments. 

From the simulated data, we can determine $Q = Q(x=0,\tau = \tau_{\rm s})$ for different $\tau_{\rm s}$. For jump-diffusion process theoretical prediction Eq.~(\ref{eq:Qdiffjump}) is valid for $\tau_{\rm s} < \frac{1}{\lambda} = T_{\rm J}$, which is the average time scale between the jumps. If the integration time step $\tau_{\rm s} > T_{\rm J}$, the jump-diffusion process behaves like diffusion process with the apparent diffusion coefficient $\widetilde{D^{(2)}}$ which has the relation, $\widetilde{D^{(2)}} \tau_{\rm s} = D^{(2)} \tau_{\rm s} + \sigma_\xi^2$. 

To prove this relation, the jump-diffusion equation can be written using Euler's method as

\begin{equation}
x(t + \tau_{\rm s}) - x(t) = D^{(1)}(x)~\tau_{\rm s} + \sqrt{D^{(2)}(x)}~\Delta W_t + \xi~\Delta J_t .
\label{eq:jumpdiffEuler}
\end{equation}

For $\tau_{\rm s} > \frac{1}{\lambda}$, $\Delta J_t = 1$ due to zero-one jump law, the Eq.~(\ref{eq:jumpdiffEuler}) becomes

\begin{equation}
x(t + \tau_{\rm s}) - x(t) = D^{(1)}(x)~\tau_{\rm s} + \sqrt{D^{(2)}(x)~\tau_s}~\eta + \xi .
\end{equation}

% \newpage
$\sqrt{D^{(2)}(x)~\tau_s}~\eta + \xi$ is the sum of two Gaussian random variables which is equivalent to $\sqrt{\widetilde{D^{(2)}}(x)~\tau_{\rm s}}~\widetilde{\eta} = \sqrt{D^{(2)}(x)~\tau_s + \sigma_\xi^2}~\widetilde{\eta}$. $\widetilde{\eta} \sim N(0,1)$ is the standard normal Gaussian white noise.

Therefore, the jump-diffusion equation for $\lambda > \frac{1}{\tau_{\rm s}}$ becomes

\begin{equation}
x(t + \tau_{\rm s}) - x(t) = D^{(1)}(x)~\tau_{\rm s} + \sqrt{\widetilde{D^{(2)}}(x)~\tau_{\rm s}}~\widetilde{\eta},
\end{equation}

respectively,

\begin{equation}
{\rm d}x = D^{(1)}(x)~{\rm d}t + \sqrt{\widetilde{D^{(2)}}(x)}~{\rm d}\widetilde{W}_t ~,
\end{equation}

where $\widetilde{D^{(2)}}$ is the apparent diffusion coefficient and  $\widetilde{W}_t$ is the Wiener process.

\subsection{$\Theta$- and $Q$-criterion for General Diffusion and Jump-Diffusion Processes}\label{app:simNLMN}

\begin{figure*}[t]
\centering
\includegraphics[width=\textwidth]{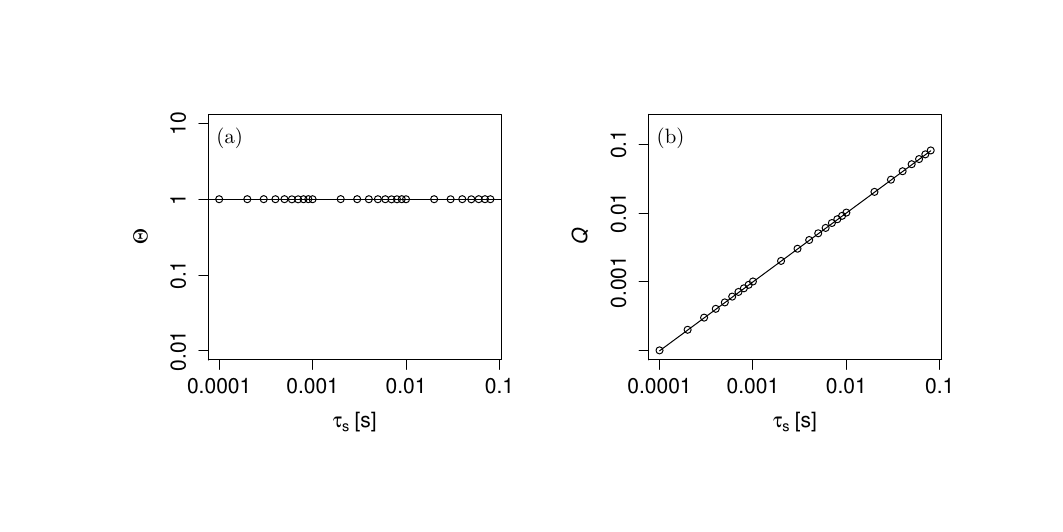}
\caption{\label{fig:diffNLMN} $\Theta$- and $Q$-criterion of diffusion process at $x = 0$ with $D^{(1)}(x) = -x^3$ and $D^{(2)}(x) = 1+x^2$, generated for different integration time steps $\tau_{\rm s}$, are plotted in double logarithmic scale. Theoretical predictions according to Eqs.~(\ref{eq:Thetadiffjump}) and (\ref{eq:Qdiffjump}) are presented with the black solid lines. Standard errors are smaller than the symbol size.}
\end{figure*}

\begin{figure*}[t]
\centering
\includegraphics[width=\textwidth]{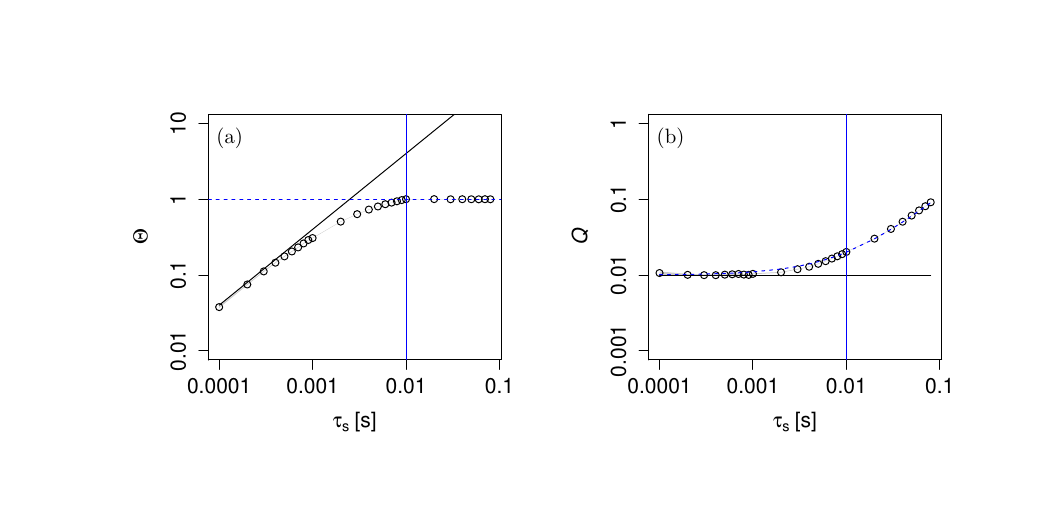}
\caption{\label{fig:JDNLMN} $\Theta$- and $Q$-criterion of jump-diffusion process at $x = 0$ with $D^{(1)}(x) = -x^3$, $D^{(2)}(x) = 1+x^2$, $\lambda = 100~{\rm s^{-1}}$ and $\sigma_\xi^2 = 1$, generated for different integration time steps $\tau_{\rm s}$, are plotted in double logarithmic scale. The vertical blue lines show the jump time scale $T_{\rm J} = \frac{1}{\lambda}$. Theoretical predictions according to Eqs.~(\ref{eq:Thetadiffjump}) and (\ref{eq:Qdiffjump}) are presented with the black solid lines. The blue dashed curve $Q = D^{(2)}(0) \tau_{\rm s} + \sigma_\xi^2$ and line $\Theta = 1$ show the behavior after $\tau_{\rm s} > T_{\rm J} = \frac{1}{\lambda}$. Standard errors are smaller than the symbol size.}
\end{figure*}

Now, the $\Theta$- and $Q$- criterion are determined for both diffusion and jump-diffusion processes with non-linear drift and multiplicative diffusion term. We numerically integrate Eq.~(\ref{eq:diffusion}) with $D^{(1)}(x) = -x^3$ and $D^{(2)}(x) = 1+x^2$, and Eq.~(\ref{eq:jumpDiffusion}) for the case with additional jump term with $\lambda = 100~{\rm s^{-1}}$ and $\sigma_\xi^2 = 1$ for different integration time step $\tau_{\rm s}$. The result of the non-linear diffusion process is shown in Fig.~\ref{fig:diffNLMN} and jump-diffusion process in Fig.~\ref{fig:JDNLMN}. 

We can observe the similar behavior in the case with linear drift and additive diffusion noise.

\section{Consequences of downsampling of the data} \label{app:downsampling}

In real world, empirical data are given for a fixed sampling time $\tau_{\rm s}$ due to experimental constraints. To handle this effect, we have to downsample the numerical data in which a new time step $\tau_{\rm ds}$ is defined for downsampling time step. For the estimation of the KM conditional moments, it is essential to see in which relation $\tau_{\rm ds}$ is with the correlation time $T_{\rm C} = \frac{1}{\gamma}$ and with the average time scale between the jumps $T_{\rm J} = \frac{1}{\lambda}$. 

Influence of downsampling on the diffusion coefficient is first examined. In order to study the downsampled approximation of the diffusion term, it is integrated over a finite downsampled time $\tau_{\rm ds}$. In the It\^o sense, the integral is generally interpreted as \cite{Hanson2007, Friedrich2011}

\begin{eqnarray}
\label{eq:diffInt}
\int_t ^{t+\tau_{\rm ds}} g(x(s),s)~{\rm d}W_s 
&=& g(x(t),t)~ \int_t ^{t+\tau_{\rm ds}} {\rm d}W_s \cr \nonumber\\
%&\stackrel{\text{ms}}{=}&  
&=& g(x(t),t) \sum_ {k=1} ^{N_\tau} (W_{k} - W_{k-1}) %\cr \nonumber\\
\end{eqnarray}

Therefore, the integral of diffusion part in (jump-)diffusion process becomes

\begin{eqnarray}
    \int_t ^{t+\tau_{\rm ds}} \sqrt{D^{(2)}(x)}~{\rm d}W_s  
    &=& \sqrt{D^{(2)}(x)} \sum_ {k=1} ^{N_\tau} \Delta W_k \cr \nonumber\\
    &=&\sqrt{D^{(2)}(x)~\tau_{\rm s}} \sum_ {k=1} ^{N_\tau} \eta_k ~. 
\end{eqnarray}

where $\eta_k \sim N(0,\sigma_{\eta}^2 = 1)$ is the standard normal distributed random variable. To estimate the summation $\sum_ {k=1} ^{N_\tau} \eta_k$, $N_\tau = \frac{\tau_{ds}}{\tau_{s}}$ is the number of time steps in the downsampled time step $\tau_{\rm ds}$. Let us define the summation as a new effective noise $\eta_{\rm eff}$ for diffusion term as,

\begin{equation}
\eta_{\rm eff} = \sum_{k=1}^{N_\tau} \eta_k 
\end{equation}

where variance of $\eta_{\rm eff}$ will be $\sigma_{\eta, {\rm eff}}^2 = N_\tau \sigma_{\eta}^2 = N_\tau = \frac{\tau_{\rm ds}}{\tau_{\rm s}}$, (as $\sigma_{\eta}^2 = 1$), and $\eta_{\rm eff}$ tends to a Gaussian random variable. Then, the integral becomes

\begin{equation}
\int_t^{t+\tau_{\rm ds}} \sqrt{D^{(2)}(x)}~{\rm d}W_s \simeq \sqrt{D^{(2)}(x)~\tau_{\rm s}}~\eta_{\rm eff} = \sqrt{D^{(2)}(x)~\tau_{\rm ds}}~
\eta\label{eq:diffeff}
\end{equation}

where $\eta \sim N(0,1)$ is the standard normal distributed. Therefore, we can conclude that diffusion coefficient remains unchanged with downsampling and summation of the noises approaches to a Gaussian white noise.

Influence of downsampling on the jump amplitude is then examined. The jump term $\xi~{\rm d}J_t$ will provide the following stochastic integral

\begin{equation}
\int_t^{t+\tau_{\rm ds}} \xi~{\rm d}J_s ~.
\end{equation}

To deal with this integral, we use the integral relation \cite{Hanson2007, Tabar2019} 

\begin{equation}
\int_t ^{t+\tau_{\rm ds}} h(x(s),s)~{\rm d}J_s \stackrel{\text{ms}}{=}  \sum_ {k=1} ^{J_t} h(x(T^-_k),T^-_k)~, 
\end{equation}

where $\stackrel{\text{ms}}{=}$ is It\^o mean square equals. The $T_k^{-}$  denotes the limit from the left to jump time $T_k$. Therefore, the jump integral becomes

\begin{equation}
\int_t ^{t+\tau_{\rm ds}} \xi~{\rm d}J_s = \sum_ {k=1} ^{J_t} \xi_k ~,
\end{equation}

where $\xi_k \sim N(0,\sigma_\xi^2)$ is the Gaussian distributed random variable with zero mean and variance $\sigma_\xi^2$. To estimate the summation $\sum_ {k=1} ^{J_t} \xi_k$, we note that the average time step of jumps in original time series is $\frac{1}{\lambda}$, then the average number of jumps becomes $J_t \simeq \lambda \tau_{\rm ds}$ in the downsampled time step $\tau_{\rm ds}$. Let us define the summation as a new noise $\xi_{\rm eff}$ for jump amplitude as,

\begin{equation}
\xi_{\rm eff} = \sum_{k=1}^{J_t} \xi_k~, 
\end{equation}

where variance of $\xi_{\rm eff}$ will be $\sigma_{\xi, {\rm eff}} ^2 \simeq \lambda \tau_{\rm ds} \sigma_{\xi} ^2$ and $\xi_{\rm eff}$ tends to a Gaussian random variable. Then, we find,

\begin{equation}
\int_t ^{t+\tau_{\rm ds}} \xi~{\rm d}J_s 
= \xi_{\rm eff}%(t) 
\simeq \sqrt{\lambda \tau_{\rm ds}}~\xi~,
\label{eq:jumpeff}
\end{equation}

where $\xi \sim N(0,\sigma_\xi^2)$ is Gaussian distributed with variance of jump amplitude $\sigma_\xi^2)$. Therefore, if we downsample with $\tau_{\rm ds} \gg \frac{1}{\lambda}$, $J_t \gg 1$, then zero-one jump law is no longer fulfilled. However, it can be realized as a jump event with the effective jump amplitude

\begin{equation}\label{eq:jumpampeff}
    \sigma_{\xi, {\rm eff}}^2 = \lambda \sigma_{\xi}^2 \tau_{\rm ds} \;.
\end{equation}

\begin{figure*}[t]
\centering
\includegraphics[width=\textwidth]{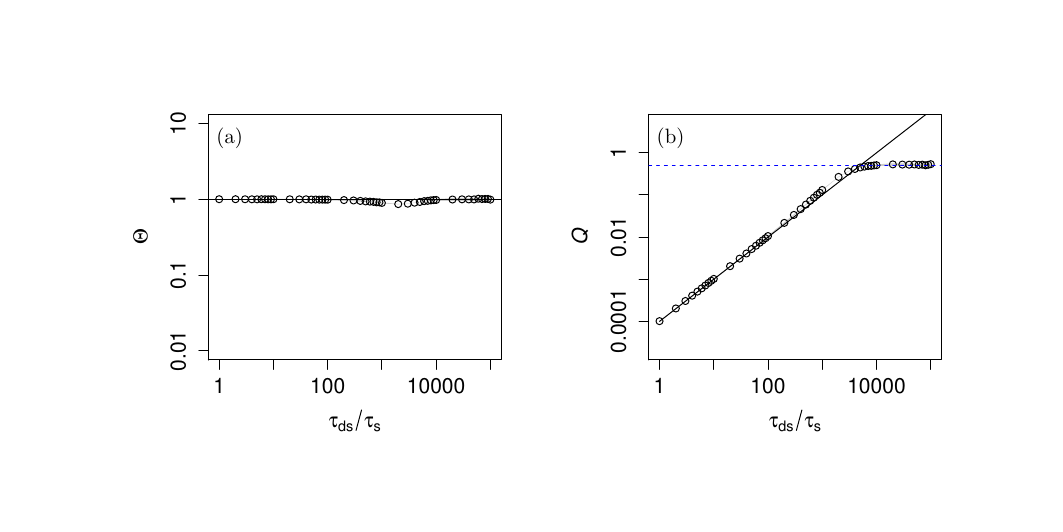}
\caption{\label{fig:diffNLMNds} $\Theta$- and $Q$-criterion of downsampled diffusion process at $x = 0$ with $D^{(1)}(x) = -x^3$ and $D^{(2)}(x) = 1+x^2$ are plotted in double logarithmic scale. Theoretical predictions according to Eqs.~(\ref{eq:Thetadiffjump}) and (\ref{eq:Qdiffjump}) are presented with the black solid lines. The blue dashed line in (a) shows $Q = \sigma_x^2$. Standard errors are smaller than the symbol size.}
\end{figure*}

\begin{figure*}[t]
\centering
\includegraphics[width=\textwidth]{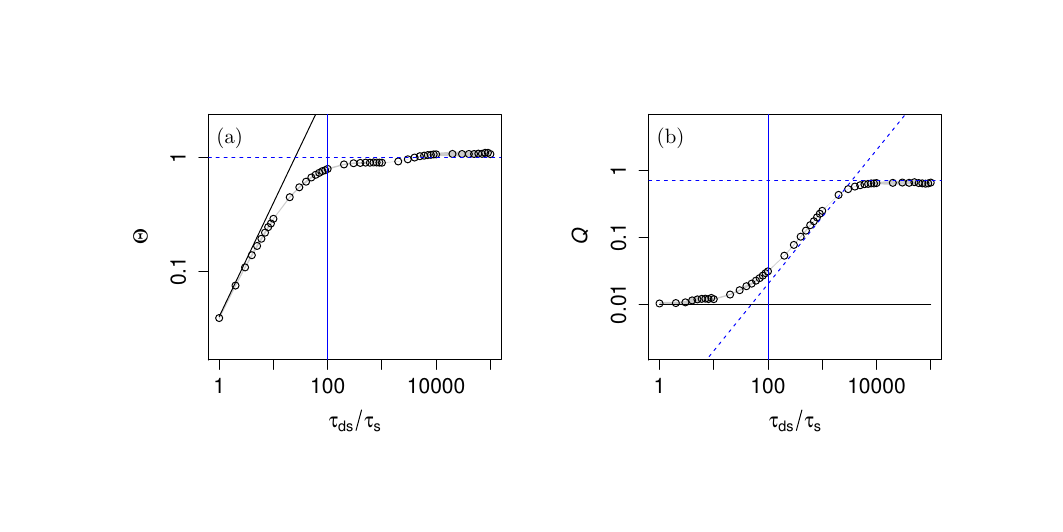}
\caption{\label{fig:JDNLMNds} $\Theta$- and $Q$-criterion of downsampled OU jump-diffusion process at $x = 0$ with $D^{(1)}(x) = -x^3$, $D^{(2)}(x) = 1+x^2$, $\lambda = 100~{\rm s^{-1}}$ and $\sigma_\xi^2 = 1$ are plotted in double logarithmic scale. The vertical blue lines show the jump time scale $T_{\rm J} = \frac{1}{\lambda}$. Theoretical predictions according to Eqs.~\ref{eq:Thetadiffjump}) and (\ref{eq:Qdiffjump}) are presented with the black solid lines. The blue dashed lines show $Q = (D^{(2)}(0) + \lambda \sigma_\xi^2)\tau_{\rm ds}$ for $\tau_{\rm ds} > T_{\rm J}$ and $Q = \sigma_x^2$ (horizontal line), and $\Theta = \Lambda = 1$. Standard errors are smaller than the symbol size.}
\end{figure*}

In addition, the $\Theta$- and $Q$-criterion are determined for both downsampled diffusion and jump-diffusion processes with non-linear drift and multiplicative diffusion terms. We numerically integrate Eq.~(\ref{eq:diffusion}) with $D^{(1)}(x) = -x^3$ and $D^{(2)}(x) = 1+x^2$, and Eq.~(\ref{eq:jumpDiffusion}) for the case with additional jump term with $\lambda = 100~{\rm s^{-1}}$ and $\sigma_\xi^2 = 1$ with a fixed sampling time $\tau_{\rm s} = 10^{-4}~\mathrm{s}$. The results are shown in Fig.~\ref{fig:diffNLMNds} for diffusion process, and in Fig.~\ref{fig:JDNLMNds} for jump-diffusion process.

We can observe the similar behavior in the case with linear drift and additive diffusion noise.

So far for the jump diffusion processes, different cases of between the time scales $T_{\rm J}$ and $T_{\rm C}$ are analyzed. It remains to study how the magnitude of jump noise with respect to the diffusion noise affect the stochastic process.

\section{Ornstein-Uhlenbeck process in the presence of jumps}

In the presence of compound Poisson jump process $Z_t = \xi J_t$ with constant jump amplitude $\sigma_\xi^2$ and jump rate $\lambda$, Ornstein-Uhlenbeck (OU) Process $x = x(t)$ can be written as Eq.~(\ref{eq:OUjump}). 

By introducing the substituting variable $y = x \mathrm{e}^{\gamma t}$, applying the It\^o product rule, 

\begin{equation}
    \mathrm{d}(f \cdot g) = f \cdot \mathrm{d}g + g \cdot \mathrm{d}f +\mathrm{d}f \cdot \mathrm{d}g ,
    \label{eq:ItoProduct}
\end{equation}

and using the following properties, 

\begin{equation}
    (\textrm{d}t)^i~(\textrm{d}W_t)^j  = 0, \textrm{for}~ i, j \geq 1,
    \label{eq:dtdW}
\end{equation}

and

\begin{equation}
    (\textrm{d}t)^i~(\textrm{d}J_t)^j  = 0, \textrm{for}~ i, j \geq 1,
    \label{eq:dtdJ}
\end{equation}

we can solve Eq.~(\ref{eq:OUjump}). The solution in $\textrm{d}t$~precision is

\begin{equation}
x = x_0 \mathrm{e}^{-\gamma t} + \sqrt{D} \int_0^t \mathrm{e}^{-\gamma (t - t^\prime)} \mathrm{d}W_{t^\prime} + \int_0^t \mathrm{e}^{-\gamma (t - t^\prime)} \mathrm{d}Z_{t^\prime},%\xi \mathrm{d}J_{t^\prime},
\label{eq:OUjumpsol}
\end{equation}

where $x_0 = x(0)$.

The mean of the OU process can be calculated as

\begin{equation}
\langle x \rangle = x_0 \mathrm{e}^{-\gamma t}.  
\label{eq:OUjumpmean}
\end{equation}

and $\langle x \rangle = 0$ for sufficiently long simulation time $t \to \infty$.

The covariance of OU process with jump can be calculated using the following properties, 

\begin{equation}
    (\textrm{d}W_t)^i~(\textrm{d}J_t)^j  = 0, \textrm{for}~ i, j \geq 1,
    \label{eq:dWdJ}
\end{equation}

\begin{equation}
    \textrm{Cov}(\textrm{d}W_s, \textrm{d}W_t) = \langle \textrm{d}W_s~\textrm{d}W_t \rangle = \delta(s-t)~\textrm{d}s~\textrm{d}t,
    \label{eq:WienerIncCov}
\end{equation}

and 

\begin{equation}
    \textrm{Cov}(\textrm{d}Z_s, \textrm{d}Z_t) = \langle \textrm{d}Z_s~\textrm{d}Z_t \rangle = \lambda \sigma_\xi^2~\delta(s-t)~\textrm{d}s~\textrm{d}t,
    \label{eq:CompoundPoissonIncCov}
\end{equation}

such that

\begin{equation}
C_{xx}(t,\tau) = \mathrm{Cov}(x(t) x(t+\tau)) = \frac{D + \lambda \sigma_\xi^2}{2 \gamma} \left[ \mathrm{e}^{-\gamma \tau} - \mathrm{e}^{-\gamma(\tau + 2t)} \right].
\label{eq:OUjumpCov}
\end{equation}

We can calculate the variance of the OU process with jump from $C_{xx}(t,\tau = 0)$ which is

\begin{equation}
\mathrm{Var}(x) = \langle x^2 - \langle x \rangle^2 \rangle = \frac{D + \lambda \sigma_\xi^2}{2 \gamma}~\mathrm{e}^{-2t}~.
\label{eq:OUjumpVar}
\end{equation}

For sufficiently long simulation time $t \to \infty$, the covariance becomes

\begin{equation}
C_{xx}(\tau) = \frac{D + \lambda \sigma_\xi^2}{2 \gamma}~\mathrm{e}^{-\gamma \tau}~,
\label{eq:OUjumpCov2}
\end{equation}

and the variance 
\begin{equation}
\mathrm{Var}(x) = \frac{D + \lambda \sigma_\xi^2}{2 \gamma}~. 
\label{eq:OUjumpVar2}
\end{equation}

\begin{figure}[t]
\centering
\includegraphics[width=\textwidth]{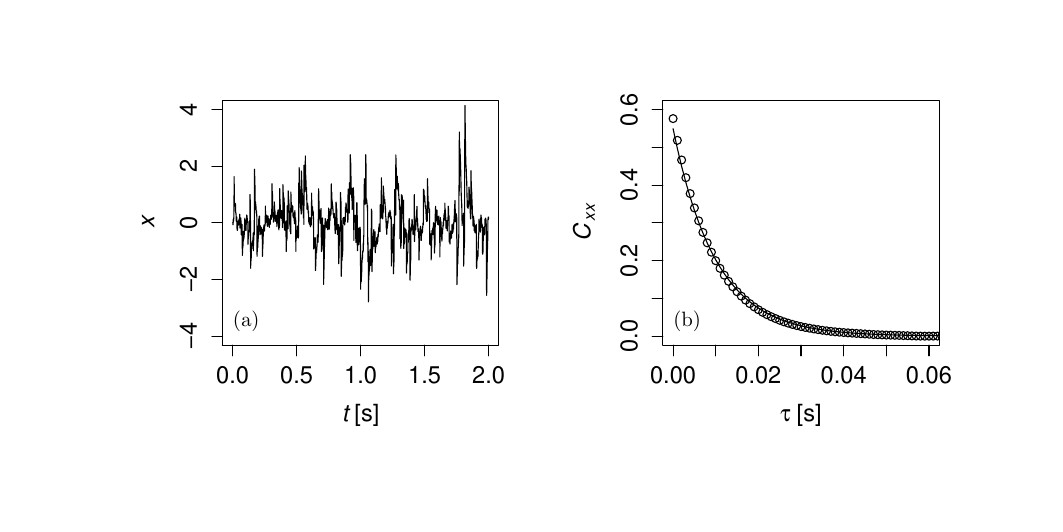}
\caption{Timeseries of OU jump-diffusion processes, (a), and its covariance function $C_{xx}(\tau)$, (b). The evaluated $C_{xx}(\tau)$ is plotted in open circles and its theoretical result from Eq.~(\ref{eq:OUjumpCov2}) are plotted with solid line.}
\label{fig:OUjump}
\end{figure}

The detailed discussions of the properties of stochastic processes that are used in this derivation can be seen in Refs.~\cite{Lin2023t, Tabar2019, Hanson2007, Gardiner1985}. In order to verify it, a synthetic time series of the OU jump-diffusion process is generated for $\Delta t = 10^{-3}~{\rm s}$ with $\gamma = 100~{\rm s^{-1}}$, $D = 10~{\rm s^{-1}}$ and for the additional jump terms with $\lambda = 100~{\rm s^{-1}}$ and $\sigma_\xi^2 = 1$. In Fig.~\ref{fig:OUjump} the timeseries of the OU jump-diffusion processes is shown. Its covariance function $C_{xx}(\tau)$ is also evaluated and plotted together with the theoretical result from Eq.~(\ref{eq:OUjumpCov2}). Here, we can observed that the correlation time $T_{\rm C}$ depends solely on $\gamma$ but not on the jump rate $\lambda$.

\section*{Data and Code Availability}
The data and codes are available upon request.

\section*{References}
\bibliographystyle{iopart-num} 
\bibliography{ref}

\end{document}